\documentclass[aps,prl,reprint,superscriptaddress,amsmath,amssymb,floatfix]{revtex4-2}

\usepackage[T1]{fontenc}
\usepackage{lmodern}
\usepackage{amsmath,amssymb,bm,mathtools}
\usepackage{graphicx}
\usepackage{microtype}
\usepackage[colorlinks=true,linkcolor=blue,citecolor=blue,urlcolor=blue]{hyperref}
\hypersetup{pdftitle={Non-Hermitian Skin Effect from Radiative Coupling in a Reciprocal Chiral Medium},pdfauthor={Kin Hung Fung, Changhao Meng, Yixin Xiao, and C. T. Chan}}

\newcommand{\ii}{\mathrm{i}}
\newcommand{\ee}{\mathrm{e}}
\newcommand{\br}{\beta}
\newcommand{\Om}{\Omega}
\newcommand{\sig}{\sigma}
\newcommand{\etazero}{\eta_0}
\newcommand{\GammaNR}{\Gamma_{\rm nr}}
\newcommand{\Omeff}{\Omega_{\rm eff}}
\newcommand{\GEE}{\mathbf{G}_{EE}}
\newcommand{\zhat}{\hat{\mathbf z}}
\newcommand{\RR}{\mathbf R}

\begin{document}

\title{Non-Hermitian Skin Effect from Radiative Coupling in a Reciprocal Chiral Medium}

\author{Kin Hung Fung}
\email{funguiuc@gmail.com}
\affiliation{Department of Physics, The Hong Kong University of Science and Technology, Hong Kong, China}
\author{Changhao Meng}
\affiliation{Department of Physics, The Hong Kong University of Science and Technology, Hong Kong, China}
\author{Yixin Xiao}
\affiliation{Department of Physics, The Hong Kong University of Science and Technology, Hong Kong, China}
\author{C. T. Chan}
\affiliation{Department of Physics, The Hong Kong University of Science and Technology, Hong Kong, China}
\affiliation{Institute for Advanced Study, The Hong Kong University of Science and Technology, Hong Kong, China}

\begin{abstract}
We show that long-range radiative coupling in a passive chiral medium produces an unusual non-Hermitian skin effect in a chain of dipoles.  The skin-mode profile consists of a long-range light-line branch-cut contribution, a non-exponential factor $1/[n(\ln n)^2]$ with $n$ being the site number.  Finite-chain calculations show that part of the open-boundary spectrum contracts toward the Bloch spectrum, whereas the number of boundary-localized modes remains extensive.  The retarded radiative coupling makes both the real-space profile and the finite-size spectral signatures of boundary localization depart from standard finite-range and pure exponential coupling theory.
\end{abstract}

\maketitle

\textit{Introduction.}---The non-Hermitian skin effect (NHSE) is characterized by the boundary accumulation of an extensive number of bulk eigenmodes under open boundary conditions.  For one-dimensional lattices with finite-range coupling, non-Bloch band theory usually replaces the unit Bloch circle $\beta=\ee^{\ii ka}$ by a generalized Brillouin zone (GBZ) in the complex-$\beta$ plane \cite{HatanoNelson1996,YaoWang2018,Kunst2018,YokomizoMurakami2019,Okuma2020,Bergholtz2021,WangZhongFan2024}.  The equal-modulus condition for a finite-order characteristic equation then determines both the open-boundary spectrum and the exponential envelope $|\beta|^n$ in the thermodynamic limit ($N\to \infty$).  This mechanism has been realized in photonic, acoustic, mechanical, electrical, and atomic systems through asymmetric couplings, controlled dissipation, feedback, and synthetic gauge fields \cite{Weidemann2020,Xiao2020,Helbig2020,Brandenbourger2019,Zhang2021Acoustic,Liang2022}.  Reciprocal systems with internal polarization or pseudospin degrees of freedom can also support paired skin effects, with symmetry-related modes localized at opposite boundaries \cite{Hofmann2020,Fang2022GDSE,Wang2023GDSE,WangYouJen2022,Han2026Helical}.  Radiative NHSE has also been reported in semiconductor photonic-crystal slabs \cite{TalukderPaiella2025}. 

A central spectral signature of the NHSE is sensitivity to boundary conditions.  For a finite-range chain, the large-$N$ spectrum under open boundary conditions (OBCs) generally differs from the Bloch spectrum under periodic boundary conditions (PBCs) and is instead determined by the GBZ.  Retarded radiative coupling makes this comparison more subtle.  Recent work on long-range emitter chains showed that even the PBC spectrum of a finite ring need not coincide with the Bloch spectrum of the infinite chain~\cite{SturmKressPalffy2026}, while critical and mesoscopic skin effects can exhibit strong spectral size dependence~\cite{Li2020Critical,YokomizoMurakami2021Scaling,Poddubny2024Mesoscopic}.  The conventional finite-range expectation of a size-independent OBC spectrum separated from the Bloch spectrum therefore cannot be assumed.  One must instead determine how the finite-chain OBC spectrum evolves relative to the infinite-chain Bloch spectrum and whether boundary localization persists as their spectral difference changes.

\begin{figure}[b]
\centering
\includegraphics[width=0.97\columnwidth]{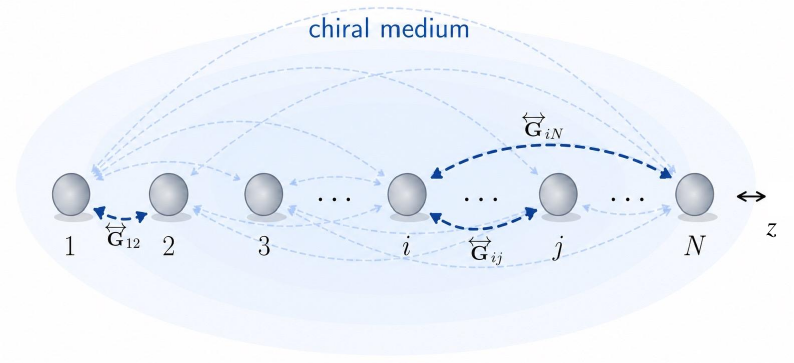}
\caption{Schematic of a chain of long-range coupled electric dipoles in a reciprocal chiral medium.  The full retarded electromagnetic Green tensor $\overleftrightarrow{G}_{ij}$ couples every pair of sites ($i$ and $j$).  Reciprocity relates the two transverse circular-polarization subspaces, and produces equal-frequency skin pairs whose spatial profiles are related by reversal of the site index.}
\label{fig:schematic_prl}
\end{figure}

To address this question, we study a straight chain of electric dipoles in a passive reciprocal Pasteur (chiral) medium [Figure~\ref{fig:schematic_prl}].  Its two transverse eigenchannels are exactly circularly polarized, and their skin modes localize at opposite boundaries.  Within a fixed channel, the leading right- and left-going fields have the wavenumbers of the two helicity eigenmodes of the host, whose complex values differ in both real and imaginary parts.  Each transverse circular-polarization subspace consequently acquires a directional phase and attenuation bias.  Reciprocity reverses the bias for the opposite circular polarization and produces equal-frequency mode pairs localized at opposite boundaries.  

\textit{Coupled dipoles in a reciprocal chiral medium.}---The physical system is formulated as a coupled-dipole theory in a simple reciprocal chiral medium as follows. With time dependence $\ee^{-\ii\omega t}$ for given frequency $\omega$, the reciprocal chiral constitutive relations are
\begin{equation}
 \mathbf D=\epsilon_0\epsilon_r\mathbf E+\ii\frac{\kappa}{c}\mathbf H,
 \qquad
 \mathbf B=\mu_0\mu_r\mathbf H-\ii\frac{\kappa}{c}\mathbf E,
 \label{eq:constitutive_prl}
\end{equation}
where $\mathbf{D}$ is the electric displacement field, $\mathbf{E}$ is the electric field, $\mathbf{B}$ is the magnetic flux density, $\mathbf{H}$ is the magnetic field, $\epsilon_0$ and $\mu_0$ are the permittivity and permeability of free space, $\epsilon_r$ and $\mu_r$ are the relative permittivity and permeability, $\kappa$ is the chirality parameter, and $c$ is the speed of light in vacuum.  The dipoles satisfy
\begin{equation}
 \alpha_{\rm rad}^{-1}\mathbf p_i-
 \sum_{j\ne i}\GEE[(i-j)a\zhat]\mathbf p_j=0,
 \label{eq:cde_prl}
\end{equation}
where the electric Green tensor obeys
\begin{multline}
 \left[\nabla\!\times\!\nabla\!\times-2\frac{\omega\kappa}{c}\nabla\!\times
 -\frac{\omega^2}{c^2}(n_h^2-\kappa^2)\right]\GEE(\RR)\\
 =\omega^2\mu_0\mu_r\,\delta(\RR)\mathbf I ,
 \label{eq:wave_prl}
\end{multline}
where $n_h=\sqrt{\epsilon_r\mu_r}$.  On the chain axis, axial rotational symmetry makes the transverse response diagonal in the circular basis $p_\sig=(p_x+\ii\sig p_y)/\sqrt2$.  To link our equations to the standard NHSE models, we consider the case of weakly dispersive $\kappa$ at a reference frequency $\omega_0$ (the single-particle resonance frequency).  The off-diagonal matrix elements in Eq.~(\ref{eq:cde_prl}) are then \cite{SM}
\begin{align}
G_\sig^R(m)&=g_{-\sig}\ee^{\ii x_{-\sig}m}
 \left(\frac{x_{-\sig}^2}{m}+\frac{\ii x_{-\sig}}{m^2}-\frac{1}{2m^3}\right)
 -\frac{g_\sig\ee^{\ii x_\sig m}}{2m^3},\nonumber\\[-2pt]
G_\sig^L(m)&=g_\sig\ee^{\ii x_\sig m}
 \left(\frac{x_\sig^2}{m}+\frac{\ii x_\sig}{m^2}-\frac{1}{2m^3}\right)
 -\frac{g_{-\sig}\ee^{\ii x_{-\sig}m}}{2m^3},
\label{eq:directional_prl}
\end{align}
where $x_\sig=\etazero(n_h+\sig\kappa)$, $\etazero=\omega_0a/c$, $x_0=(x_++x_-)/2$, and $g_\sig$ is the helicity-dependent coupling coefficient incorporating the resonator-strength normalization and the host response. The leading $1/m$ terms give
\begin{equation}
 G_\sig^{R,L}(m)\sim \frac{C_\sig^{R,L}}{m}
 \ee^{\ii x_0 m}\ee^{\mp\ii\sig(x_+-x_-) m/2}.
 \label{eq:gauge_prl}
\end{equation}
The real part of $(x_+-x_-)$ produces opposite phase shifts for the two propagation directions, whereas the imaginary part produces opposite attenuation shifts.  Equation~(\ref{eq:gauge_prl}) therefore shows that the leading radiative coupling is equivalent to a polarization-dependent complex shift of the Bloch wave number; its imaginary part is the familiar imaginary gauge field.  Reversing $\sig$ reverses the shift and exchanges the two boundaries, as required by reciprocity.  The full Maxwell interaction contains additional polarization-dependent prefactors and the $m^{-2}$ and $m^{-3}$ near-field terms in Eq.~(\ref{eq:directional_prl}).  These terms are retained in all spectra and eigenvectors, while Eq.~(\ref{eq:gauge_prl}) isolates the physical origin of the directional bias.

Evaluating the Green tensor and the radiative self-energy at the reference frequency $\omega_0$ gives a linearized coupled-dipole eigenvalue problem.  We denote its complex eigenvalue by $\Om=\omega/\omega_0$.  The two circular-polarization subspaces have identical eigenvalues, and their right eigenvectors are related by reversal of the site index.  The Supplemental Material extends the calculation to the nonlinear outgoing-wave eigenvalue problem, with the constitutive parameters and retarded Green tensor evaluated self-consistently at complex frequency.  The resulting modes preserve the exact reciprocal pairing and display the strongest dispersive shifts near the light lines \cite{SM}.

\begin{figure}[!t]
\centering
\includegraphics[width=\columnwidth]{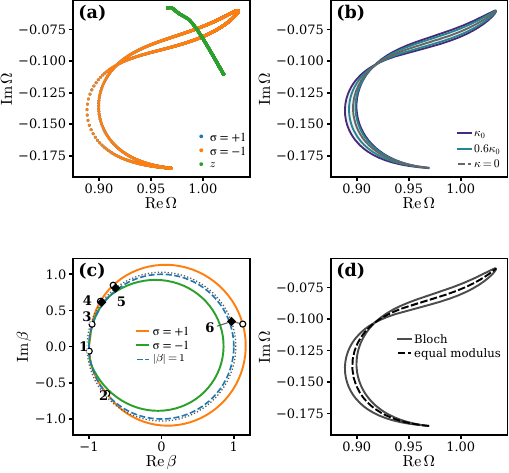}
\caption{Bloch spectrum and analytically continued equal-modulus locus for $\epsilon_r=1.3221+0.020\ii$, $\mu_r=1+0.002\ii$, baseline chirality $\kappa_0=0.080+0.004\ii$, $\etazero=\pi$, and $\GammaNR=1/120$.  The lattice constant $a$ is chosen to be half of the corresponding free-space resonance wavelength of the dipole resonators.  (a) Bloch spectra of the two reciprocal transverse circular-polarization subspaces and of the longitudinal control.  (b) Transverse Bloch spectra for $\kappa=\kappa_0$, $0.6\kappa_0$, and $0$; for $\kappa=0$, the spectral loop collapses to a reciprocal arc.  (c) Directly computed solutions of the analytically continued equal-modulus condition in the complex-$\beta$ plane.  The dashed curve is the unit Bloch circle; the other circles mark the boundaries of the convergence annulus, and the symbols mark the light-line branch points.  The $\sig=-1$ solutions are obtained from the $\sig=+1$ solutions by $\beta\mapsto\beta^{-1}$.  The numbered points correspond to modes 1--6 in Fig.~\ref{fig:obc_skin_prl}(b).  For modes 4--6, open circles mark the equal-modulus solutions and filled diamonds the $\beta$ values fitted from the eigenvector envelopes at $N=1200$.  (d) Bloch spectrum $\mathcal B$ and analytically continued equal-modulus locus $\mathcal C$ in the complex-$\Omega$ plane.  The plotted loci use the directly calculated points.}
\label{fig:bulk_fourpanel}
\end{figure}

\begin{figure}[!t]
\centering
\includegraphics[width=\columnwidth]{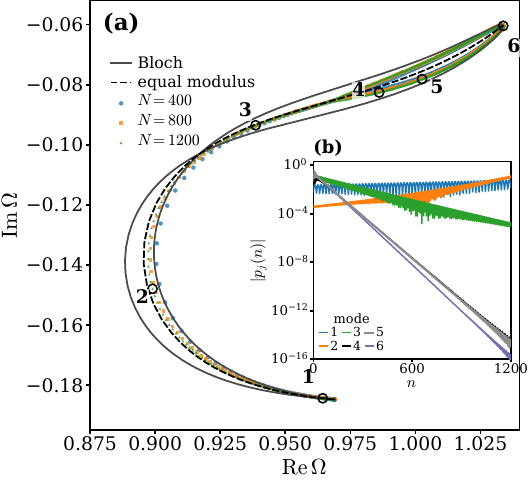}
\caption{Open-boundary spectra and reciprocal boundary localization.  (a) Eigenvalues of the linearized coupled-dipole matrix for open chains with $N=400$, $800$, and $1200$, plotted together with the Bloch spectrum $\mathcal B$ and the analytically continued equal-modulus locus $\mathcal C$.  (b) Representative right-eigenvector amplitudes for the $\sig=-1$ subspace at $N=1200$.  The $\sig=+1$ partners have reversed site profiles.  The examples span weak two-boundary localization (mode 1), broad opposite-edge profiles (modes 2 and 3), and light-line-limited strong decay (modes 4--6).}
\label{fig:obc_skin_prl}
\end{figure}

To connect the reciprocal boundary pairing in Fig.~\ref{fig:schematic_prl} with its spectral origin, we first examine the Bloch and complex-$\beta$ structure in Fig.~\ref{fig:bulk_fourpanel}, before comparing it with the open-chain spectra and profiles in Fig.~\ref{fig:obc_skin_prl}.  For a periodic chain, the scalar Bloch symbol of the transverse circular-polarization subspace $\sig$ is
\begin{equation}
 \Omega_\sig(\br)=\Omeff-
 \sum_{m\ge1}\left[G_\sig^R(m)\br^m+G_\sig^L(m)\br^{-m}\right].
 \label{eq:disp_prl}
\end{equation}
where $\Omeff=1-\ii\GammaNR-\Sigma^{(0)}_{\rm rad}$ contains the nonradiative linewidth $\GammaNR$ and the radiative self-energy $\Sigma^{(0)}_{\rm rad}$ \cite{SM}.  The series resums exactly to
\begin{align}
 \Omega_\sig(\br)&=\Omeff-g_{-\sig}\Lambda(x_{-\sig},\br)-g_\sig\Lambda(x_\sig,\br^{-1}),
 \nonumber\\
 \Lambda(x,z)&=x^2\operatorname{Li}_1(z\ee^{\ii x})+\ii x\operatorname{Li}_2(z\ee^{\ii x})
 \nonumber\\
 &\quad-\tfrac12\bigl[\operatorname{Li}_3(z\ee^{\ii x})+\operatorname{Li}_3(z^{-1}\ee^{\ii x})\bigr].
 \label{eq:closed_prl}
\end{align}
Under periodic boundary conditions, $\br=\ee^{\ii ka}$ lies on the unit circle and Eq.~(\ref{eq:disp_prl}) gives the Bloch spectrum.  Reciprocity implies $\Omega_\sig(\br)=\Omega_{-\sig}(\br^{-1})$, manifest in Eq.~(\ref{eq:closed_prl}).  The lattice sums contain $\operatorname{Li}_1(\br\ee^{\ii x_\sig})=-\log[1-\br\ee^{\ii x_\sig}]$ and the corresponding functions of $\br^{-1}$.  These logarithms generate light-line branch points.  The original real-space series is analytic only inside the convergence annulus
\begin{equation}
 \mathcal A=\{\br:\ee^{-\delta_-}<|\br|<\ee^{\delta_-}\},
 \qquad
 \delta_\sig=\operatorname{Im}x_\sig,
 \label{eq:annulus_prl}
\end{equation}
where $\delta_-=\min_\sig\delta_\sig$ for the parameters used here.  Inside $\mathcal A$, the polylogarithmic expression coincides with the convergent real-space lattice sum.  In contrast to a finite-range lattice, the characteristic function is branch valued rather than polynomial.  Its non-Bloch structure is therefore governed jointly by isolated zeros and branch cuts.  Range truncation replaces this structure by a cutoff-dependent polynomial whose additional roots accumulate near the boundaries of $\mathcal A$ \cite{SM}.  

\begin{figure}[!t]
\centering
\includegraphics[width=\columnwidth]{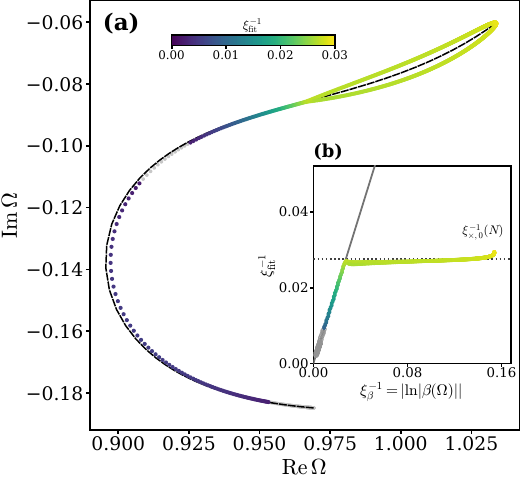}
\caption{Mode-by-mode comparison of spatial decay rates for the $1080$ localized modes at $N=1200$.  (a) Effective inverse decay rates obtained by fitting the eigenvector envelopes over sites 150--900.  (b) Comparison with the complex-$\beta$-pole prediction $\xi_\beta^{-1}=|\ln|\beta||$ from the nearest point of the analytically continued equal-modulus locus.  The two rates agree for roots near the unit circle.  For more strongly localized roots, the fitted rates approach the finite-size crossover value in Eq.~(\ref{eq:crossover_prl}) because the light-line branch-cut contribution limits the observable decay.}
\label{fig:decay_rates}
\end{figure}

Before we illustrate the long-range effects on finite chains, here we use Figure~\ref{fig:bulk_fourpanel} to summarize the usual Bloch spectrum and non-Bloch equal-modulus solutions of the complex $\beta$.  The two transverse circular-polarization solutions trace the same closed Bloch loop with opposite orientations, as required by reciprocity, whereas the longitudinal control remains an open arc [Fig.~\ref{fig:bulk_fourpanel}(a)].  Reducing $\kappa$ continuously narrows the transverse loop, which collapses to the arc for $\kappa=0$ [Fig.~\ref{fig:bulk_fourpanel}(b)].  This indicates that chirality is the origin of the point-gap opening.  Figure~\ref{fig:bulk_fourpanel}(c) shows the analytically continued equal-modulus solutions of $\beta$.  The two curves, related by $\beta\mapsto\beta^{-1}$, leave the unit circle and approach the convergence boundaries.  Unlike in finite-range non-Bloch theory, however, the equal-modulus construction is not sufficient here: it identifies the isolated complex-$\beta$ poles but does not by itself determine the finite-chain decay.  Near the convergence boundaries, the light-line branch cuts limit the observable localization rate, causing the decay rate $|\ln|\beta||$ to overestimate the fitted decay rate of strongly localized modes [Figs.~\ref{fig:obc_skin_prl}(b) and~\ref{fig:decay_rates}].  In the complex-$\Omega$ plane, the equal-modulus locus $\mathcal C$ is displaced from $\mathcal B$ except near their intersections [Fig.~\ref{fig:bulk_fourpanel}(d)].  The resulting finite-chain spectra and profiles are shown in Fig.~\ref{fig:obc_skin_prl}.

\textit{Finite-size open-boundary spectrum and eigenvector localization.}---Figure~\ref{fig:obc_skin_prl}(a) shows nonuniform evolution from $N=400$ to $1200$: part of the OBC spectrum contracts toward the Bloch spectrum $\mathcal B$, whereas the most displaced eigenvalues remain near the equal-modulus locus $\mathcal C$; both are plotted in Fig.~\ref{fig:bulk_fourpanel}(d).

The analytically continued equal-modulus locus $\mathcal C$ follows the complex-$\beta$ branches associated with the most displaced finite-chain eigenvalues.  Using the nearest-locus criterion defined in the Supplemental Material, the associated fraction decreases with increasing chain length.  The locus therefore describes the finite-size spectral branches that remain most strongly displaced from the Bloch spectrum. We will see that this analytical continued solutions are not the true thermodynamic OBC solutions in the radiative coupling cases here.

The six numbered complex-$\beta$ solutions in Fig.~\ref{fig:bulk_fourpanel}(c) correspond to the real-space profiles in Fig.~\ref{fig:obc_skin_prl}(b), which separate into three decay types.  Mode 1 is weakly localized and influenced by both ends.  Modes 2 and 3 are broad, $\beta$-pole-controlled profiles at opposite boundaries: mode 2 grows rightward, while mode 3 has an oscillatory left-edge decay.  Modes 4 and 5 share a faster, light-line-limited left-edge decay, with mode 6 the limiting upper-tip case.  Such localization is extensive: the count of modes with edge imbalance $|W_R-W_L|>1/2$, where $W_L$ and $W_R$ are the eigenvector weights in the outermost tenth of sites at the two edges, rises from $159$ of $200$ modes to $1080$ of $1200$ modes in the $\sig=+1$ subspace.  The reciprocal $\sig=-1$ profiles occupy the opposite edges, while longitudinal modes remain extended \cite{SM}.

To quantify the partial contraction seen in Fig.~\ref{fig:obc_skin_prl}(a), we use two simple diagnostics: the mean distance of the OBC eigenvalues from the Bloch spectrum and the fraction lying outside a fixed spectral neighborhood.  Both decrease with $N$.  Forward and reverse set distances give the same trend \cite{SM}.

The contrast between the $\beta$-pole-controlled modes 1--3 and the light-line-limited modes 4--6 in Fig.~\ref{fig:obc_skin_prl}(b) is tested over all localized modes in Fig.~\ref{fig:decay_rates}.  In Fig.~\ref{fig:bulk_fourpanel}(c), the equal-modulus solutions of modes 4--6 (open circles) lie beyond the convergence annulus, and their decay rate $|\ln|\beta||$ overestimates the $\beta$ values fitted from the eigenvector envelopes (filled diamonds).

\textit{Boundary profile from $\beta$-poles and light-line branch cuts.}---To interpret this crossover, we express the boundary solution as a generating function of the complex variable $\beta$.  Its singularities have two different origins.  Isolated zeros of the analytically continued denominator produce exponential terms, whereas the logarithmic light-line branch points produce branch cuts.  Deforming the coefficient contour separates these contributions explicitly.  For a boundary excitation with nonzero overlap with the branch point of the less attenuated helicity channel, the amplitude at a distance $n$ from the occupied boundary has the asymptotic form \cite{SM}
\begin{equation}
 p_n=A_p\ee^{-\xi_\beta^{-1}n+\ii q_p n}
 +A_{\rm LL}\frac{\ee^{-\delta_-n+\ii q_{\rm LL}n}}
 {n(\ln n)^2}\,[1+o(1)].
 \label{eq:polecut_prl}
\end{equation}
In a driven nanoparticle chain, radiation damping can induce an exponential-to-algebraic crossover in plasmon propagation \cite{BrandstetterKunc2016}, while non-Hermitian systems in two or more dimensions can exhibit algebraic skin localization \cite{ZhangShuSun2025}; Eq.~(\ref{eq:polecut_prl}) instead describes a quasi-one-dimensional skin eigenmode with an attenuated light-line contribution proportional to $\ee^{-\delta_-n}/[n(\ln n)^2]$.
Here $\xi_\beta^{-1}=|\ln|\beta_p||$ and $q_p$ are the decay rate and momentum of the isolated pole $\beta_p$, and $q_{\rm LL}$ is the momentum of the nearest branch point.  The isolated complex-$\beta$ pole produces the exponential profile observed over an intermediate spatial range.  The light-line branch cut produces the second term: the attenuation constant $\delta_-$ of the least-attenuated radiative channel fixes its exponential factor, while the logarithmic singularity supplies the non-exponential factor $1/[n(\ln n)^2]$.  For generic nonzero light-line overlap, the asymptotic exponential decay rate is therefore $\min(\xi_\beta^{-1},\delta_-)$.  The branch-cut term controls when $\xi_\beta^{-1}>\delta_-$, so propagation attenuation sets an asymptotic ceiling on the localization rate.  Equating the two contributions near the far boundary of a chain of length $N$ gives the finite-size crossover rate
\begin{equation}
 \xi_\times^{-1}(N)=\delta_-+
 \frac{\ln N+2\ln\ln N+\chi_A}{N}
 +O[(N\ln N)^{-1}],
 \label{eq:crossover_prl}
\end{equation}
where $\chi_A=\ln|A_p/A_{\rm LL}|$ contains the ratio of the $\beta$ pole and branch-cut amplitudes.  Eq.~(\ref{eq:crossover_prl}) gives the finite-size crossover rate approached by strongly localized modes under a single-exponential fit and connects that rate directly to the light-line attenuation.

When $\xi_\beta^{-1}$ is below $0.02$, the median ratio of the fitted rate to $\xi_\beta^{-1}$ lies between $0.98$ and $1.01$.  For larger values of $\xi_\beta^{-1}$, the fitted rates cross over to a narrow plateau around $0.0277$, in quantitative agreement with Eq.~(\ref{eq:crossover_prl}) evaluated with $\chi_A=0$, while $\xi_\beta^{-1}$ reaches $0.154$.  The fitted crossover is insensitive to the spatial fitting interval \cite{SM}.

The leading $1/m$ far-field expression reproduces the fitted decay rates of representative strongly localized modes to within $0.3\%$, confirming that the directional radiative term controls their dominant envelope.  All spectra and eigenvectors reported here are calculated with the full retarded kernel, including the $m^{-1}$, $m^{-2}$, and $m^{-3}$ contributions.

\begin{figure}[!t]
\centering
\includegraphics[width=\columnwidth]{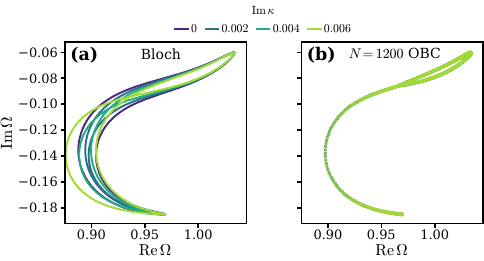}
\caption{Dependence on the dissipative part of the chirality parameter, $\operatorname{Im}\kappa$.  (a) Bloch spectra of the two transverse circular-polarization subspaces.  (b) Open-boundary spectra for $N=1200$.  Varying $\operatorname{Im}\kappa$ strongly deforms the Bloch spectrum by moving a light-line branch point, whereas the OBC spectrum remains nearly unchanged over the same range.  The more visible open-chain response occurs in the directional decay of the eigenvectors.  All material parameters in the sweep satisfy the passivity conditions derived in the Supplemental Material \cite{SM}.}
\label{fig:kappa_compact}
\end{figure}

Figure~\ref{fig:kappa_compact} shows that the finite-chain OBC spectrum is nearly insensitive to $\operatorname{Im}\kappa$ over the range studied, even though the Bloch spectrum changes strongly.  At fixed $\operatorname{Im}n_h$, varying $\operatorname{Im}\kappa$ shifts $\delta_\pm=\etazero[\operatorname{Im}n_h\pm\operatorname{Im}\kappa]$ oppositely while leaving their mean unchanged.  To leading far-field order, this is a diagonal similarity transformation of the open-chain matrix, which changes the eigenvector envelopes but not the eigenvalues; the weak residual motion comes from the $\kappa$-dependent prefactors and subleading near-field terms \cite{SM}.  PBCs instead fix $|\beta|=1$, so the same gauge change strongly deforms the Bloch loop.  Thus, $\operatorname{Im}\kappa$ can tune localization without appreciably shifting the open-chain resonances.  When $\kappa=0$, the two helicity propagation constants become equal and the transverse spectral loop collapses to the reciprocal arc shown in Fig.~\ref{fig:bulk_fourpanel}(b).  Including both transverse polarization subspaces restores the full reciprocal pairing: every state localized at one edge has a partner at the same complex frequency whose site profile is reversed and therefore localized at the other edge.

\textit{Conclusion.}---We have shown that long-range radiative coupling through a passive reciprocal chiral medium produces an unusual non-Hermitian skin effect in a chain of dipoles.  The retarded interaction determines the skin-mode profile through an isolated complex-$\beta$ pole and logarithmic light-line branch points.  For generic nonzero light-line overlap, attenuation in the least-attenuated radiative channel caps the asymptotic exponential decay rate, while the logarithmic singularity supplies the universal non-exponential factor $1/[n(\ln n)^2]$.  This analytic structure explains the crossover of the spatial decay observed in finite chains. Finite-chain calculations further show that part of the OBC spectrum contracts toward the Bloch spectrum as the chain length increases, whereas the number of boundary-localized modes remains extensive.  Thus, this partial spectral contraction coexists with reciprocal skin localization rather than eliminating it.  Light-line branch singularities generated by retarded radiative coupling, not only isolated complex poles, can control the asymptotic real-space form of boundary eigenstates.  In this radiative setting, the observed partial contraction toward the Bloch spectrum does not by itself imply the loss of boundary accumulation. 

\begin{acknowledgments}
We thank Sen Lin and Prof.~Z.~Q. Zhang for useful discussions.  This work is supported by RGC Hong Kong (JLFS/P-603/24).
\end{acknowledgments}

\begingroup
\small
\makeatletter
\let\auto@bib@innerbib\@empty
\makeatother

\endgroup
\makeatletter
\global\let\auto@bib\@empty
\global\let\write@bibliographystyle\relax
\makeatother

\end{document}